\documentclass{article} 
\usepackage{amsmath,amsfonts,amssymb}
\usepackage{bm} 
\usepackage{cases}
\usepackage{float}
\usepackage{mwe}
\usepackage{mathtools}
\usepackage{comment}
\usepackage{empheq}
\usepackage{booktabs}
\usepackage{textcomp}
\usepackage{algorithm}
\usepackage{algorithmic}
\usepackage{array}
\usepackage{subcaption}
\usepackage{graphicx}
\usepackage{url}
\usepackage{caption}
\usepackage{systeme}
\usepackage{adjustbox}
\usepackage{comment}
\usepackage{soul}
\usepackage{xcolor}

\usepackage{tikz}
\usetikzlibrary{positioning}
\usetikzlibrary{3d}
\usetikzlibrary{matrix}
\usetikzlibrary{decorations.text}
\usetikzlibrary{spy}

\newcommand{\steplength}{\eta}
\newcommand{\seqpenalty}[1]{\mu_{#1}}
\newcommand{\increasingfact}{\gamma}
\newcommand{\downfact}{d}

\usepackage{enumitem}

\usepackage[roman]{written}

\usepackage{graphicx}
\usepackage{amsmath} 

\usepackage[margin=2.5cm]{geometry}

\usepackage[capitalize]{cleveref}

\makeatletter
\let\@fnsymbol\@arabic
\makeatother

\begin{document}

\title{RELD: Regularization by Latent Diffusion Models for Image Restoration}

\author{
    Pasquale~Cascarano\footnotemark[1]\thanks{Department of the Arts, University of Bologna, Bologna, 40123 Italy ({\{pasquale.cascarano2, gustavo.marfia\}@unibo.it})}
    \and Lorenzo~Stacchio\footnotemark[2]\thanks{ Department of Political Sciences, Communication and International Relations, University of Macerata, Macerata, 62100, Italy ({lorenzo.stacchio@unimc.it})}
    \and Andrea Sebastiani\footnotemark[3]\thanks{ Department of Physics, Informatics and Mathematics, University of Modena and Reggio Emilia, Modena, 41125, Italy ({andrea.sebastiani@unimore.it})}
    \and Alessandro Benfenati\footnotemark[4]\thanks{ Department of Environmental and Science Policy, University of Milan, Milan, 20133, Italy ({alessandro.benfenati@unimi.it})}
    \and Ulugbek~S.~Kamilov\footnotemark[5]\thanks{ Department of Computer Science \& Engineering and Department of Electrical \& System Engineering, Washington University in St. Louis, St. Louis, MO 63130 USA ({kamilov@wustl.edu})}
    \and Gustavo~Marfia\footnotemark[1]
    }

\date{}
\maketitle

\begin{abstract}
In recent years, Diffusion Models have become the new state-of-the-art in deep generative modeling, ending the long-time dominance of Generative Adversarial Networks. Inspired by the Regularization by Denoising principle, we introduce an approach that integrates a Latent Diffusion Model, trained for the denoising task, into a variational framework using Half-Quadratic Splitting, exploiting its regularization properties.
This approach, under appropriate conditions that can be easily met in various imaging applications, allows for reduced computational cost while achieving high-quality results. The proposed strategy, called Regularization by Latent Denoising (RELD), is then tested on a dataset of natural images, for image denoising, deblurring, and super-resolution tasks. The numerical experiments show that RELD is competitive with other state-of-the-art methods, particularly achieving remarkable results when evaluated using perceptual quality metrics.
\end{abstract}

{\bf Keywords:} Diffusion Models, Latent Space, Inverse Problems, Image Restoration.\\

\section{Introduction}

Image Restoration (IR) aims at reconstructing an image $\vx \in \mathbb{R}^{n}$ from an observed data $\vb \in \mathbb{R}^{m}$. 
This encompasses tasks such as image super-resolution, deblurring, and denoising which can be casted as linear inverse problems:
\begin{equation}\label{eq:inverse-problem}
 \vb = \vA \vx + \veta,
\end{equation}
where $\vA \in \mathbb{R}^{m \times n}$ is the known measurement operator, and $\veta \in \mathbb{R}^{m}$ represents an additive random noise component with a standard deviation $\sigma_{\veta}$. 
Inverse problems are often ill-posed, meaning their solutions may not be either defined, unique or stable. This is due to their sensitivity to data perturbations, making direct inversion of 
$\vA$ challenging for obtaining accurate and reliable  solutions \cite{bertero2021introduction}.

A common approach is to provide an approximate solution $\vx^{\ast}$ of \eqref{eq:inverse-problem} by minimizing the objective function:

\begin{equation} \label{eq:reg_mod}
\vx^{\ast} \in \underset{\vx \in \mathbb{R}^{n}}{\text{argmin}} \ \ell (\vx;\vb) + \lambda \rho(\vx),
\end{equation}
which includes the data-fidelity term $\ell$, the regularization functional $\rho$, and the regularization weighting parameter $\lambda$. The data-fidelity term $\ell(\vx;\vb)$ is chosen depending on the noise perturbing the given data. In particular, for Additive White Gaussian Noise (AWGN) $\ell(\vx;\vb)=\frac{1}{2}\lVert \vA \vx - \vb \rVert^{2}_{2}$. 
The regularization parameter $\lambda >0$ is typically hand-tuned to achieve optimal restorations.

A traditional approach customizes handcrafted regularization functional $\rho(\vx)$ based on specific characteristics of the images. 
Some examples include the well-known Tikhonov regularization~\cite{golub1997tikhonov} or the Total Variation functional~\cite{rudin1994total,cascarano2021combining} and their extensions \cite{bredies2014recovering}.
Despite their solid mathematical foundation, such regularizers struggle to capture the complexity of image features thus providing suboptimal solutions \cite{arridge2019solving}.

Nowadays, data-driven regularization has outperformed handcrafted regularizers in imaging inverse problems. 
For instance, end-to-end approaches leverage highly expressive neural networks to learn a direct mapping from $\vb$ to $\vx$. More specifically, a dataset comprising diverse measurements along with their respective ground truths is constructed and a neural network is trained to minimize the empirical risk \cite{arridge2019solving}.
These methods bypass the need for the forward degradation model \eqref{eq:inverse-problem}, thus being preferable for those imaging tasks where the underlying physical processes are either uncertain or challenging to formalize through analytical expressions. 
However, these learning strategies strongly rely on large training datasets, which could be a limitation for those applications having limited access to such amount of data \cite{piening2024learning}. Furthermore, the inference process suffers from instability, since any deviations from the distribution of training set, due to noise in the measurements, out of distribution data or different acquisition models, can lead to the generation of misleading artifacts \cite{gottschling2020troublesome,antun2020instabilities,szegedy2013intriguing}. 

Other learning-based strategies aim to mitigate the strong dependence on training data by preserving the variational structure \eqref{eq:reg_mod} and learning a data-driven regularizer $\rho$ (or a function related to $\rho$). 
A popular technique belonging to this class is Plug-and-Play (PnP)  \cite{venkatakrishnan2013plug,Pendu23,cascarano2022plug,kamilov2017plug,hurault2023convergent}. 
The seminal idea of PnP  \cite{venkatakrishnan2013plug} exploits the modularity of iterative optimization algorithms to replace the proximal operator of the regularizer with any off-the-shelf denoiser \cite{kamilov2023plug}. 
Various denoisers have been considered in the PnP framework, ranging from BM3D or NLM filters \cite{venkatakrishnan2013plug} to novel learning based strategies such as CNNs. \cite{chen2016trainable,zhang2017beyond,zhang2017learning,zhang2021plug}. 
Focusing on the iterative algorithms adopted within this framework, several alternatives have been considered, including the Half-Quadratic Splitting (HQS) algorithm \cite{zhang2021plug,zhang2017learning}, the Alternating Direction Method of Multipliers (ADMM) \cite{venkatakrishnan2013plug,sreehari2016plug} and the  Fast Iterative Shrinkage-Thresholding Algorithm (FISTA) \cite{kamilov2023plug,kamilov2017plug}. 
Despite the promising empirical results obtained in many applications, the replacement of the proximal map with a denoiser undermines the correspondence with an objective function that has to be minimized \cite{hurault2023convergent}, thus limiting the subsequent convergence analysis. 
Nonetheless, in \cite{sreehari2016plug} the authors have proved a denoiser is a proximal mapping whenever it is the sub-gradient of a convex function and it is non-expansive. 
However, these two conditions are difficult to be proved and they are also rarely met by popular denoisers used in the literature \cite{reehorst2018regularization}.
Various fixed-point convergence results have been proposed in the literature under weaker conditions \cite{chan2016plug,sreehari2016plug,ryu2019plug,terris2020building}: these results require specific properties on the denoisers which reduce their representational power \cite{ryu2019plug}. 
Regularization by denoising (RED) \cite{Romano17} addresses the theoretical analysis of PnP methods by using the denoiser to define an explicit regularization functional. Nevertheless, practical challenges remain, particularly to ensure that denoisers align properly with the manifold of natural images and fulfill restrictive assumptions \cite{cohen2021regularization,hurault2023convergent}. 

Another approach leverages generative deep learning models to better approximate the image manifold to capture the full complexity of data distributions \cite{bora2017compressed,shah2018solving,pan2021exploiting,duff2024regularising}. Various architectures have been employed \cite{bond2021deep} ranging from Variational Autoencoders (VAEs) \cite{kingma2013auto}, Generative Adversarial Networks (GANs) \cite{goodfellow2014generative} to autoregressive models \cite{van2016pixel} and normalizing flows \cite{rezende2015variational}. 
More in detail, a model $\vN:\mathbb{R}^z\mapsto\mathbb{R}^n$ is trained to map a latent variable $\vz\in \mathbb{R}^z$, with $z\ll n$, from a low-dimensional space to the image space. 
This model, which is usually referred to as \emph{generator}, is used to replace the image and to reformulate the minimization problem \eqref{eq:reg_mod} into the latent space, deriving the following optimization problem:
\begin{equation} \label{eq:reg_generative_prior}
\vz^{*} = \underset{\vz \in \mathbb{R}^{n}}{\text{argmin}} \ \ell (\vN(\vz);\vb) \qquad \vx^{\ast}=\vN(\vz^{*}).
\end{equation}
In some cases, the pre-trained generator is further fine-tuned, minimizing the previous problem also in the space of model's weights \cite{pan2021exploiting,asim2020blind}, similarly to what is done within the Deep Image Prior framework \cite{ulyanov2018deep}. 

\textit{Contribution:} Motivated by the recent developments  of powerful generative models such as Diffusion Models (DMs), we introduce the novel Regularization by Latent Diffusion (RELD) algorithm, which uses Latent Diffusion Models (LDMs) to implicitly define regularizers for solving imaging inverse problems.
Inspired by the regularization properties of denoising schemes in the PnP  and RED frameworks, the LDM is trained to solve the task of image denoising. 
In particular, during the training the reverse diffusion process is conditioned on the output of a universal encoder applied to noisy natural images. 
The result of this process is then passed through a universal decoder, before being compared with the corresponding ground truth images. 
Once trained, the reverse diffusion process is utilized to construct a generative map, which we integrate as a generative prior within the variational model \eqref{eq:reg_generative_prior}, treating both the latent variable and the conditioning as variables. 
The resulting optimization problem is then solved through the iterative HQS algorithm,  dividing it into two sub-problems, which can be efficiently addressed under certain conditions: the first has a closed-form solution, while the second is approximately solved with a single gradient step, eliminating the need for a more complex iterative algorithm. 
The proposed approach is finally tested on several IR tasks such as denoising, deblurring and super-resolution and the restored images are compared with several state-of-the-art methods with respect to quantitative and qualitative metrics.

The paper is organized as follows. Section \ref{sec:2} reviews the current literature concerning the use of latent diffusion models in the field of image restoration. Section \ref{sec:3} illustrates the novel RELD model and the implemented optimization scheme. In Section \ref{sec:4} we describe all the implementation details and we report the numerical results of RELD compared to other state-of-the-art methods. Section \ref{sec:concl} presents the conclusions.

\section{Related work \label{sec:2}}

In this section, we briefly review DMs, outlining their characteristics and illustrating their state-of-the-art performances on IR tasks.

\subsection{Background on diffusion models \label{sec:background_DM}}

DMs \cite{croitoru2023diffusion} are a family of probabilistic generative models that transform the challenging and unstable generation task into a series of independent and stable reverse steps using Markov Chain modeling. 
A DM consists of a forward and a reverse process. The forward process incrementally degrades the data by adding random noise until it becomes Gaussian noise. 
The reverse process then works to remove the noise in order to generate new data samples.  
Nowdays, DMs have emerged as the new state-of-the-art family of deep generative models for the task of image synthesis, breaking the long-time dominance of GANs \cite{dhariwal2021diffusion,nichol2021improved},  in a variety of domains, like natural language processing and temporal data modeling~\cite{yang2022diffusion,croitoru2023diffusion}.
The existing literature categorize DMs as score-based Stochastic Differential
Equation (SDE) \cite{song2020score} and Denoising Diffusion Probabilistic Models (DDPMs) \cite{ho2020denoising}.

In score-based SDE models the continuous forward process is usually modeled through the SDE: 
\begin{equation} \label{eq:SDE_forward}
    dx = f(x, t)dt + g(t)dw,
\end{equation}
where $w$ is the standard Wiener process, $f(\cdot, t)$ and $g(\cdot)$ are the vector-valued drift and the scalar diffusion coefficients of $x(t)$, respectively.
The diffusion process described by the SDE in \eqref{eq:SDE_forward} can be reversed in time:
\begin{equation}\label{eq:SDE_backward}
    dx = [f(x, t) - g(t)^{2} \nabla_{x}\log p_{t}(x)]dt + g(t)dw,
\end{equation}
where $p_{t}(x)$ is the marginal probability density at timestep $t$, and the only unknown part $\nabla_{x}\log p_{t}(x)$ can be modeled as the so-called score function $s_{\theta}(x,t):=\nabla_{x}\log p_{t}(x)$ with score matching methods \cite{hyvarinen2005estimation,song2019generative}. Here, $\theta$
represents the parameters of a neural network trained to approximate the score function. We can generate data samples according to \eqref{eq:SDE_backward} by evaluating the score function $s_{\theta}(x,t)$ at each intermediate timestep during sampling, even if the initial state is Gaussian noise. Under specific hypothesis it has been proved that a well-trained denoising score function is also an ideal Gaussian denoiser. 
Concerning DDPMs it has been proved that they are discretizations of SDE. 
For the specific choice of $f(x,t) = -1/2 \beta_t x$  and $g(x, t) = \sqrt{\beta_t}$ we have the forward and reverse SDEs as the continuous version of the diffusion process in DDPM \cite{ho2020denoising}.
One forward step of DDPM is:
\begin{equation}
    x_{t}  = \sqrt{1-\beta_{t}} x_{t-1} + \sqrt{\beta_{t}}\epsilon_{t-1},
\end{equation}
with $\epsilon_{t-1} \sim \mathcal{N}(0,\vI)$. The sample $x_{t}$ is obtained by adding i.i.d. Gaussian noise with variance $\beta_{t}$ and scaling $x_{t-1}$ with $\sqrt{1-\beta_{t}}$. 
We can also sample $x_{t}$ at an arbitrary timestep $t$ from $x_{0}$: 
\begin{equation}\label{eq:x0xt}
    x_{t}  = \sqrt{\overline{\alpha}_{t}} x_{0} + \sqrt{1-\overline{\alpha}_{t}}\epsilon,
\end{equation}
where $\alpha_{t} = 1 - \beta_{t}$, $\overline{\alpha_{t}}=\prod^{t}_{s=1} \alpha_{s}$ and $\epsilon \sim \mathcal{N}(0,\vI)$. The corresponding reverse step of DDPM reads \cite{ho2020denoising}: 
\begin{equation}\label{eq:reverse_DDPM}
    x_{t-1} = \dfrac{1}{\sqrt{\alpha_{t}}} \left(x_{t} - \dfrac{\beta_{t}}{\sqrt{1 - \overline{\alpha}_{t}}} \epsilon_{\theta}(x_{t},t)\right) + \sqrt{\beta_{t}}\epsilon_{t},
\end{equation}
where $\epsilon_{\theta}(x,t)$ is a neural network, parametrized by $\theta$, designed to predict the total noise $\epsilon$ between $x_{t}$ and $x_{0}$ in \eqref{eq:x0xt}. To enable deterministic sampling, the authors in \cite{ho2020denoising} introduce Denoising Diffusion Implicit Models (DDIM). In the reverse process, DDIM utilizes the same equation as \eqref{eq:reverse_DDPM}, but without adding random noise $\epsilon_{t}$.   

\subsection{Latent diffusion models}

The first DDPMs operate directly in the high-dimensional image space. Despite its efficiency, this approach often leads to a significant computational workload because training and evaluating the associated deep neural network require repeated function evaluations and gradient computations ~\cite{ho2020denoising,rombach2022high}.  To make training and inference faster and more efficient, a new class of DM called Latent Diffusion Models (LDMs) has been introduced \cite{rombach2022high}.

LDMs separate training into two distinct phases: (a) an autoencoder is trained to provide a lower-dimensional representational space that is perceptually equivalent to the data space and (b) a denoising network is trained on the learned latent space. 
A key advantage of this approach is that the learned autoencoder could be reused to train multiple LDMs on different tasks \cite{rombach2022high}.
At the same time, this approach enables conditional image generation: it is possible to condition the reverse process of the underlying LDM using both images or texts.

Differently from DDPMs, the LDM is designed to learn a data distribution $p(z)$ defined in the latent space by gradually denoising a normally distributed variable, which corresponds to learning the reverse process of a fixed Markov Chain of length $T$.
In this approach, perceptual encoder $\mathcal{E}$ and decoder $\mathcal{D}$ models are introduced. This modifies the process by moving the optimization of the denoising network to a low-dimensional latent space,  where high-frequency details of the data are abstracted away. 
Operating in this space provides two main advantages: (a) it allows the model to focus on capturing semantic information, and (b) it reduces computational complexity, enabling a more efficient training in a lower-dimensional space.
More specifically, the denoising network $\epsilon_{\theta}(z_t, t)$, parametrized by $\theta$, is trained to predict the noise component in the latent variable $z_t$. Here, $t$ is a time step uniformly sampled from ${1, \dots, T}$.
The training objective for the LDM is to minimize the loss:
\begin{equation}
LDM := \mathbb{E}_{\mathcal{E}(x),\epsilon \sim \mathcal{N}(0, 1), t} \Big[ || \epsilon - \epsilon_{\theta}(z_t, t)||^{2}_{2} \Big],
\end{equation}
where $z_t$ is the latent representation of the data at time step $t$ perturbed by Gaussian noise according to the forward process.
The neural backbone $\epsilon_{\theta}(z_t, t)$ is usually implemented as a time-conditional U-Net \cite{ronneberger2015u,rombach2022high}, as often adopted in the literature for DDPMs~\cite{ho2020denoising,dhariwal2021diffusion}.
At inference time, since the forward process is fixed, $z_t$ can be efficiently obtained from $\mathcal{E}$ during training, and samples from $p(z)$ can be decoded into the image space through $\mathcal{D}$.

LDMs can then be reformulated to steer their generation with a conditional approach. It is indeed possible to model the conditional distributions of the form $p(z|y)$ where $y$ allows control of the synthesis
process through inputs such as text or images~\cite{rombach2022high}. The conditional generation could be implemented by exploiting an input from a different modality or domain, using concatenation or cross-attention mechanisms of its latent representation~\cite{rombach2022high}.
The first modality appears to be more effective for structural generation, while the second one is well suited for style injection~\cite{rombach2022high}.
Using conditional mechanisms, the LDM's training loss can be reformulated as follows:
\begin{equation}
LDM := \mathbb{E}_{\mathcal{E}(x),y,\epsilon \sim \mathcal{N}(0, 1), t} \Big[ || \epsilon - \epsilon_{\theta}(z_t, t, \tau_{\theta}(y) )||^{2}_{2} \Big]
\end{equation}
where $\tau_{\theta}(y)$ is a neural network that projects the conditioning input $y$ from its data space to the latent one. More specifically, $\tau_{\theta}$ could amount to the encoder $\mathcal{E}$ or a different trained model based on the conditional input $y$.

\subsection{Diffusion models for image restoration \label{sec:IIC}}

Approaches to IR predominantly rely on image-space DMs, although LDMs have also been explored \cite{li2023diffusion}. 
They can be categorized into supervised and zero-shot methods. Supervised approaches involve training DMs based on the specific task, thus facing practical challenges in acquiring large paired dataset. In contrast, zero-shot methods leverage the generative capabilities of DMs as prior without relying on extensive paired training data.
In this context, DMs have emerged as a powerful paradigm, due to their ability to capture complex data distributions while avoiding the common drawbacks of other generative models, such as mode collapse~\cite{kossale2022mode,xia2023diffir}.

In the literature, several strategies have been proposed to guide the generation process toward solutions that align the given measurements.
Overall, these methods interleave iterative steps to move toward the data manifold and iterative steps to move toward the set of feasible solutions through hard and soft consistency techniques.
Hard consistency techniques enforce strict alignment with measurements by projecting solutions onto the feasible set or minimizing constraints exactly \cite{song2023solving,kadkhodaie2021stochastic,chung2022improving}. 
Soft consistency techniques guide the solution iteratively using gradient-based updates, penalization via loss functions, or incorporating priors, allowing incremental alignment with measurements \cite{fei2023generative,luo2023refusion,xia2023diffir,chung2022diffusion}. 
These approaches are often combined to balance accuracy and computational efficiency \cite{garber2024image,chung2023prompt,zhu2023denoising,fei2023generative}.
For instance, ReSample \cite{song2023solving} develops a two-stage process leveraging LDMs and hard data consistency to provide solutions aligned with the given measurements. Diffusion Posterior Sampling (DPS) \cite{chung2022diffusion} makes use of image-based DDPMs to address both linear and nonlinear noisy inverse problems. DPS leverages manifold-constrained gradients to approximate posterior sampling, avoiding the noise amplification issues commonly encountered in projection-based methods \cite{chung2022improving}.
Similarly, in \cite{li2024decoupled} the authors adopt diffusion purification steps as prior enforcement. 
In order to deal with insufficient manifold and measurement feasibility,  DMPlug \cite{wang2024dmplug} presents a framework that redefines the reverse diffusion process as a function, ensuring both measurement and manifold feasibility through global optimization.
Efficiency has been a critical area of improvement. 
For instance, in \cite{yue2024efficient} the authors model residual shifts between high- and low-quality images, shortening the Markov Chain in the reverse diffusion process. WaveDM \cite{huang2024wavedm} leverages wavelet-based decomposition to model image distribution in frequency bands, drastically reducing computational overhead and achieving high-quality restoration with few sampling steps.
Similarly, DiffIR \cite{xia2023diffir} introduces a prior representation that guides a dynamic IR transformer, significantly reducing the number of sampling steps. 
Recently, DMs have been incorporated into existing frameworks, demonstrating their versatility. For example, PnP paradigms have also benefited from diffusion models \cite{xu2024provably, zhu2023denoising}. The initial exploration of DMs as generative denoisers within the PnP framework was presented in \cite{zhu2023denoising}, where the authors introduced DiffPIR.

Unlike other approaches, the proposed RELD derives from a variational formulation that integrates LDMs. The resulting iterative scheme utilizes the HQS method, which requires us to address at each iteration two subproblems, that can be solved efficiently. By operating entirely in the latent space, RELD reduces the computational cost while maintaining high-quality image restorations.
Inspired by the regularization properties of denoising schemes in the PnP and RED frameworks, RELD employs an LDM specifically trained for denoising tasks, using it as a conditioned generator. Furthermore, this is combined with strategies such as a warm-start initialization, a concatenation-based conditioning approach, and optimization performed not only on the latent variable but also on the concatenated conditioning. These elements work together to ensure robust alignment with the data manifold and accurate image restoration.
\section{Method \label{sec:3}}

\subsection{RELD framework}

The starting point to derive our approach is the following optimization problem:
\begin{equation}\label{eq:model}
    \argmin{\vv \in \mathbb{R}^{s}} \dfrac{1}{2}\lVert \vA\,\vN(\vv) - \vb \rVert_{2}^{2},
\end{equation}
where $\vN:=\mathcal{D}\circ \Pi \circ \vS^p$  such that 
\begin{itemize}
    \item $\vS^{p}:\rd^s\rightarrow\rd^{s}$ denotes the execution of  \cref{al:diffuse_step} for  $p$-steps;
    \item $\Pi:\rd^s\to\rd^{s_2}$ denotes the projection of a vector in $\rd^s$ into $\rd^{s_2}$, formally defined as $\Pi([\va,\vz])=\vz$ $\forall \va\in\rd^{s_1}, \vz\in\rd^{s_2}$, with $s=s_1+s_2$;
    \item $\mathcal{D}:\rd^{s_2}\to\rd^{n}$ represents the trained universal decoder described in \cite{rombach2022high}.    
\end{itemize}
In \cref{al:diffuse_step}, $\vU:\rd^s\rightarrow\rd^{s_2}$ denotes a time-conditioned U-Net. The integer $p >0$ refers to the number of denoising steps in the reverse diffusion process.  Steps 3-5 represent the updates of the standard DDIM reverse diffusion process. Step 6 shows the concatenation strategy adopted in our approach.

\begin{algorithm}[htbp]
\caption{Definition of $\vS^{p}$}
\label{al:diffuse_step}
\begin{algorithmic}[1]\small
\STATE{Set $\vv^{p} = [\va,\vz^{p}]$, with $\va\in\rd^{s_1}$, $\vz^{p} \sim \mathcal{N}(0,1)$.}
\FOR{$i=p,\ldots,1$}
\STATE $\hat{\vs}=\vU(\vv^i)$
\STATE $\hat{\vz}=\frac{1}{\sqrt{\overline{\alpha}_i}}\left(\vz^i-\sqrt{1-\overline{\alpha}_i}\hat{\vs}\right) $
\STATE $\vz^{i-1} = \sqrt{\overline{\alpha}_{i-1}} \hat{\vz} + \sqrt{1 - \overline{\alpha}_{i-1}} \hat{\vs}$
\STATE $\vv^{i-1}=[\va, {\vz}^{i-1}]$
\ENDFOR
\STATE \textbf{Return} $\vv^0$
\end{algorithmic}
\end{algorithm}

Considering an auxiliary variable $\vt \in \mathbb{R}^{n}$, subject to the constraint $\vt=\vN(\vv)$, the problem \eqref{eq:model} reads as: 

\begin{equation}
    \argmin{\vt \in \mathbb{R}^{n}} \dfrac{1}{2} \lVert \vA \vt - \vb \rVert_{2}^{2}  \quad\text{subject to}\quad \vt = \vN(\vv).
\end{equation} 
This constrained problem can be relaxed by considering the following quadratic penalized objective function:
\begin{equation}\label{eq:augmented_Lagrangian_constrained}
    \begin{split}
    L(\vv,\vt) =&\, \dfrac{1}{2} \lVert \vA \vt - \vb \rVert_{2}^{2} + \frac{\mu}{2}\lVert \vN(\vv) - \vt \rVert_{2}^{2}. 
    \end{split}
\end{equation}
By applying the HQS method, we alternatively minimize \eqref{eq:augmented_Lagrangian_constrained} with respect to $\vt$ and $\vv$. The resulting method is described in Algorithm \ref{al:RELD} and we refer to it as RELD.

\begin{algorithm}[htbp]
\caption{RELD}
\label{al:RELD}
\begin{algorithmic}[1]\small
\STATE{Set $\vv^{0} = [\va^0,\vz^{0}]$, with $\va^0=\tau(\vb)$, $\vz^{0} \sim \mathcal{N}(0,1)$. Select $\mu_{0}, \gamma > 0$.}
\FOR{$k=1,\dots,K_{\text{MAX}}$}
\STATE $\vv^{k}\gets \vS^p(\vv^{k-1})$
\STATE $\vt^{k+1}=\argmin{\vt\in\rd^d}\frac{1}{2}\lVert\vA\vt-\vb\rVert^2+\mu_{k}\lVert\vt-\vN(\vv^k)\rVert^2$

\medskip

\STATE $\vv^{k+1} =  \vv^k -\steplength \nabla_\vv L(\vv^k,\vt^{k+1}) $
\ENDFOR
\end{algorithmic}
\end{algorithm}

Employing the HQS methods provides some advantages. Upon suitable conditions of the operator $\vA$, the update for $\vt^{k+1}$ at line 4 of \cref{al:RELD} can be computed using a closed-form expression by means of the Fast Fourier Transform \cite{zhu2023denoising}.
The gradient step at line 5 can be easily performed using automatic differentiation. We point out that the actual subproblem to be solved reads as $\vv^{k+1}=\argmin{\vv} L(\vv,\vt^{k+1})$. Although the solution is not achieved via a closed formula or via an iterative solver, the numerical experience shows that even a single gradient step is sufficient to achieve reliable results.

A further advantage of the proposed method consists in employing an already trained LDM, tailored for image denoising tasks, thus reducing the overall computational cost.  
We point out that in the first iterate of Algorithm \ref{al:RELD}, the latent variable $\vv$ is the concatenation of two parts: the first is $\tau(\vb)$, that is a function of the corrupted data $\vb$, while the second is initialized as pure random Gaussian noise. 
The choice for our method is to consider $\tau\equiv \mathcal{E}$, \emph{i.e.}, we include the embedding of $\vb$. 
This induces a warm-start on the initial iterate of the optimization process. Unlike other existing approaches discussed in Section \ref{sec:IIC}, the optimization is performed both on the latent and on the concatenated variables.

\begin{figure}[]
    \centering
    \includegraphics[width=\linewidth]{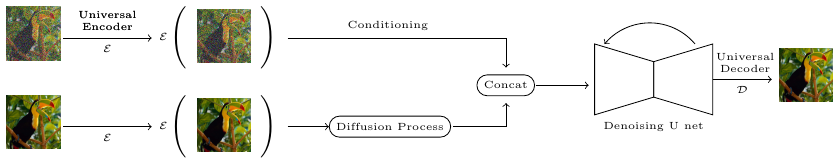}
    \caption{Denoising LDM architecture training pipeline.}
    \label{fig:architecture}
\end{figure}

Concerning the considered LDM, we remark that our approach is inspired by the regularization properties of denoising schemes used in PnP and RED frameworks.  
Therefore, we propose to optimize the LDM for image denoising and then to use it as a generative prior. 
More specifically, we proceed following the general training pipeline of an LDM. As architectures, we consider the autoencoder provided by~\cite{rombach2022high} and then we train from scratch a time-conditioned U-Net. 
The U-Net is optimized to learn to denoise an image through the concatenation conditioning modality.
In particular, the input image is projected into the latent space and perturbed through $t$ steps to become a full random Gaussian noise image. This is then concatenated with the embedding of a noisy image which by adding synthetic Gaussian noise to the original ground truth. The U-Net is then optimized to remove the noise. The architecture of the proposed system is reported in Figure~\ref{fig:architecture}.

\section{Numerical Results \label{sec:4}}

In this section, we compare the proposed RELD with different strategies for IR tasks. 
More specifically, we considered the HQS-PnP method introduced in \cite{zhang2017learning} and a specific version of RED using an ADMM scheme with an automatic regularization parameter selection rule \cite{cascarano2024constrained}. For the sake of brevity, we refer to these two methods as PnP and RED, respectively. 
For these two approaches we consider the set of CNN denoisers introduced in \cite{zhang2017learning} as priors. For both methods, the maximum number of iterations is set at 400. In addition, a stopping criterion is included, setting the threshold on relative difference of the iterates at $10^{-4}$.

Furthermore, we compare our RELD with two DM-based strategies. In particular,  we selected the Diffusion Posterior Sampling (DPS) method \cite{chung2023diffusion} and a more recent technique, named DiffPIR \cite{zhu2023denoising}, which considers a diffusion model within a PnP framework. 

\subsection{Implementation details and evaluation metrics}

Our LDM considers the time-conditioned U-Net and the universal encoder $\mathcal{E}$ and decoder $\mathcal{D}$ described in~\cite{rombach2022high}. 
However, we train the U-Net architecture from scratch on the Smartphone Image Denoising Dataset (SIDD)~\cite{abdelhamed2018high}.
This dataset consists of $\approx 30,000$ noisy images, acquired from 10 scenes under different lighting conditions, using five representative smartphone cameras, and their generated ground truth images. 
Nonetheless, we made some adjustments on this dataset to employ it in our framework. 
Indeed, the images contained in the SSID dataset are all at high resolution ($\approx 4k$), but LDMs are usually trained with a resolution ranging from $256 \times 256$ to $768 \times 768$. 
Therefore, we constructed our own dataset starting from SSID by extracting several patches from the original images corresponding to our target resolution (\emph{i.e.}, $256 \times 256$).

We further corrupted the noisy images within SIDD by adding synthetic random Gaussian noise with $\sigma$ randomly ranging from $0$ to $0.25$,  to obtain the conditioning images from which we computed the embeddings by applying the universal encoder $\mathcal{E}$.

Our model is trained over the entire dataset for a total of $60000$ steps of the Adam optimizer, with a batch size of $64$. 
For all the other hyperparameters, we followed the same protocol proposed in~\cite{rombach2022high}.

We employed images from the Set5 dataset \cite{bevilacqua2012low}, in order to evaluate the performance of the proposed procedure. 

All the experiments are executed on a server powered by NVIDIA Tesla V100 GPU with 32GB memory.

We compare the results using the Peak Signal-to-Noise-Ratio (PSNR) metric and other three standard metrics for IR tasks assessing visual quality, as the Naturalness Image Quality Evaluator (NIQE) \cite{mittal2012making}, the Perception-based Image Quality Evaluator (PIQE) \cite{venkatanath2015blind} and the Learned Perceptual Image Patch Similarity (LPIPS)  measures \cite{zhang2018unreasonable}.

\subsection{Ablation study and settings}
In this section we investigate the behaviour of RELD with respect to the choice of its hyperparameters. In particular, we determine the influence of increasing sequence of penalty parameters $\mu_k$ and the number of denoising steps $p$ performed in the reverse diffusion process. In the next experiments, the acquisition operator $\vA$ in \eqref{eq:inverse-problem} is chosen as a blurring operator, defined by its point spread function (PSF). In particular, we select a Gaussian PSF specified through its standard deviation $\sigma_{\vA}$.
\subsubsection{Penalty parameter sequence}
Concerning the sequence of penalty parameters, we adopt an increasing sequence defined as $\seqpenalty{k}:=\increasingfact^k\seqpenalty{0}$, with $\increasingfact>1$. 
In order to understand the sensibility of the model to the selection of the hyperparameters $\seqpenalty{0}$ and $\increasingfact$,  where they denote the penalty parameter and the increasing factor, respectively. 
As an example, we consider the \textit{Butterfly} image from Set5, setting $\sigma_{\vA}=1$, $\sigma_{\veta}=15$ and we inspect the influence of the choice of the starting penalty parameter $\seqpenalty{0}$ and the increasing rate $\increasingfact$, on the quality of the restored images in terms of visual inspection and the PSNR metric. 
We consider 40 samples of $\seqpenalty{0}$ in the range $[0.05, 2]$ and three values of $\increasingfact=1, 1.01,1.05$. We point out that setting $\increasingfact=1$ corresponds to not increasing the values of the sequence $\seqpenalty{k}$. The number of iterations $K_{\text{MAX}}$ is fixed at 100 and the step-length $\steplength$ is set at $10^{-3}$.
In \cref{fig:ablation_study_penalty} we report a subset of resulting images. By visual inspection, our model proves to be stable. The resulting mean LPIPS of 0.15 with a standard deviation of just 0.02 also confirms the stability of our approach over the range of values of the two hyperparameters.
\input{images/figure_table}

\subsubsection{Number of diffusion denoising step}
\begin{figure}
    \centering
    \newcommand\factor{0.15}
    \subfloat[$p=1$]{
    \begin{tikzpicture}
    \node [name=c]{	\includegraphics[width=\factor\textwidth]{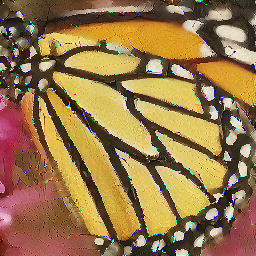}};
    \node[align=center, text=white] at (-0.35*\factor\textwidth, -0.42*\factor\textwidth) {\footnotesize 17.24};
    \end{tikzpicture}
    }
    \hspace{4mm}
    \subfloat[$p=5$]{
    \begin{tikzpicture}
    \node [name=c]{	\includegraphics[width=\factor\textwidth]{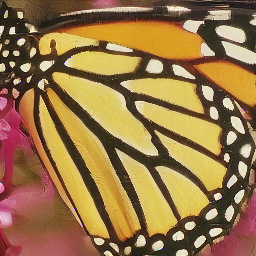}};
    \node[align=center, text=white] at (-0.35*\factor\textwidth, -0.42*\factor\textwidth) {\footnotesize 23.40};
    \end{tikzpicture}
    }
    \\
    \subfloat[$p=10$]{
    \begin{tikzpicture}
    \node [name=c]{	\includegraphics[width=\factor\textwidth]{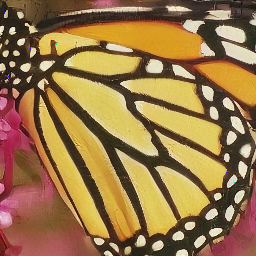}};
    \node[align=center, text=white] at (-0.35*\factor\textwidth, -0.42*\factor\textwidth) {\footnotesize 23.94};
    \end{tikzpicture}
    }
    \hspace{4mm}
    \subfloat[$p=15$]{
    \begin{tikzpicture}
    \node [name=c]{	\includegraphics[width=\factor\textwidth]{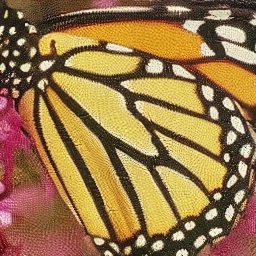}};
    \node[align=center, text=white] at (-0.35*\factor\textwidth, -0.42*\factor\textwidth) {\footnotesize 21.70};
    \end{tikzpicture}}
    \caption{Reconstruction results of RELD with different diffusion step $p$. The PSNR values are reported on top of each image.}
    \label{fig:ablation_study_step}
\end{figure}

Concerning the choice the number of diffusion steps $p$ performed using the diffusion model $\vS_{p}$, we run the algorithm selecting different values of steps $p$ ranging from 1 to 50. We can observe in \cref{fig:ablation_study_step} that the quality of the reconstructions degrades for more than 10 diffusion steps, introducing high-frequency artifacts. In fact,  the PSNR metrics computed on the images in \cref{fig:ablation_study_step} confirm the previous considerations. 

According to these experiments, we set the number of diffusion step $p=10$, the increasing factor $\increasingfact=1.01$ and initial parameter $\seqpenalty{0}=1$. Concerning the maximum number of iterations we fixed it at 100, considering the step-length $\steplength=10^{-3}$.

\subsection{Deblurring}
In this section, we focus on the task of image deblurring with AWGN. 
We simulate blurry and noisy data by applying the image formation model \eqref{eq:inverse-problem} to the images from Set5 referred to as ground truths.

The second step of our RELD method, related to the variable $\vt$, is performed computing the proximal operator in a closed form exploiting the Fast Fourier transform, assuming periodic boundary conditions \cite{hansen2006deblurring}. 

In the first set of experiments, we consider a Gaussian blur with $\sigma_{\vA}=0.7$ and $\sigma_{\veta}=35$. We report in \cref{tab:3} the averages of the quality metrics computed on the reconstructions for the images in Set5.  It is evident that our RELD outperforms the other methods in terms of perceptual quality, achieving the best scores in NIQE and PIQE and the second-best results in terms of LPIPS, closely following DPS. 

However, since latent-space optimization does not strictly enforce pixel-level fidelity, RELD may not achieve the highest PSNR. We report in \cref{fig:5} the qualitative results on an image belonging to Set5. As shown, the slight content variations introduced by RELD penalize pixel-based metrics such as PSNR. Nevertheless, we observe that all the considered competing methods tend to oversmooth the natural textures of the original image, whereas RELD provides sharper details.

\begin{table}[!h]
\centering
\caption{Values of PSNR, NIQE, PIQE and LPIPS for the images in Set5 corrupted setting $\sigma_{\vA}=0.7$ and $\sigma_{\veta}=35$. Best results are highlighted in bold. }
\begin{tabular}{|c|c|c|c|c|c|}
\hline 
& RED  & PNP & DPS & DiffPIR & RELD \\
\hline
PSNR $\uparrow$ & 27.17  & 27.36 & \textbf{27.82} & 26.71 & 26.94 \\
NIQE $\downarrow$ & 7.45  & 7.70 & 6.20 & 5.59 & \textbf{4.84} \\
PIQE $\downarrow$ & 56.48  &  65.89 & 57.41 & 30.85 & \textbf{18.15} \\
LPIPS $\downarrow$ & 0.318  &  0.272 & \textbf{0.214} & 0.241 & 0.223 \\

\hline
\end{tabular}
\label{tab:3}
\end{table}

In the second batch of experiments, we set $\sigma_{\vA}=1$ and $\sigma_{\veta}=25$. Similarly to the previous experiments, we report the average metrics on the Set5 images in \cref{tab:4}. RELD proves to have the best performances in terms of no-reference metrics PIQE and NIQE also in this setting. In addition, RELD achieves the best result in terms of LPIPS showing to be significantly competitive in terms of perceptual quality. 

As a general comment, we remark that among the diffusion-based approaches, RELD achieves the best trade-off between distortion and perceptual quality. It obtains a similar PSNR but significantly outperforms DPS and DiffPIR in perceptual metrics, which demonstrates its effectiveness in restoring natural and visually coherent images under severe degradation.

\begin{table}[!h]
\centering
\caption{Values of PSNR, NIQE, PIQE and LPIPS for the images in Set5 corrupted setting $\sigma_{\vA}=1$ and $\sigma_{\veta}=25$. Best results are highlighted in bold.}
\begin{tabular}{|c|c|c|c|c|c|}
\hline
& RED & PNP & DPS & DiffPIR & RELD \\
\hline
PSNR $\uparrow$ & 27.95 & \textbf{27.96} & 27.63  & 27.18 & 27.23 \\
NIQE $\downarrow$ & 7.88 & 8.43 & 6.83 & 5.82 & \textbf{5.35}\\
PIQE $\downarrow$ & 72.88 &  76.99 & 57.57 & 34.81 & \textbf{29.67}\\
LPIPS $\downarrow$ & 0.279 &  0.265 & 0.214 & 0.235 & \textbf{0.198}\\

\hline
\end{tabular}
\label{tab:4}
\end{table}

\input{images/figure_table2_bird_cut}
\subsection{Super-Resolution}
In this section, we focus on the super-resolution task, considering an acquisition operator $\vA$ which is the composition of a decimation operator and a blurring operator, already described in the previous set of experiments. The decimation operator takes a column/row every $\downfact$ columns/rows.
The proposed algorithm is implemented considering the same choices as in the previous section, considering the closed form expression proposed in \cite{zhao2016fast} to perform the step which updates the auxiliary variable $\vt$.

In the first experiment on the super-resolution problem, we consider $\downfact=2$, $\sigma_{\vA}=1.2$ and $\sigma_{\veta}=15$. The average results on the images in Set5 are reported in \cref{tab:5} and they demonstrate that RELD outperforms the competing methods with regard to NIQE, PIQE and LPIPS.

\begin{table}
\centering
\caption{Values of PSNR, NIQE, PIQE and LPIPS for the images in Set5 corrupted setting $\downfact=2$, $\sigma_{\vA}=1.2$ and $\sigma_{\veta}=15$. Best results are highlighted in bold. }
\begin{tabular}{|c|c|c|c|c|c|}
\hline
& RED & PNP & DPS & DiffPIR & RELD \\
\hline
PSNR $\uparrow$ & \textbf{26.40} & 26.15 & 23.63 & 23.12 & 25.72   \\
NIQE $\downarrow$ & 7.84 & 7.58 & 6.68 & 5.47 & \textbf{5.01}  \\
PIQE $\downarrow$ & 78.52 & 87.15 & 54.35 & 46.23 & \textbf{40.44}  \\
LPIPS $\downarrow$ & 0.314 & 0.323 & 0.291 & 0.285 & \textbf{0.281}  \\

\hline
\end{tabular}
\label{tab:5}
\end{table}

We perform a similar experiment increasing the downsampling factor $\downfact$, in order to test the behaviour of our model when the number of pixels in the measurements is very limited. In particular, we set $\downfact=4$, $\sigma_{\vA}=1$ and $\sigma_{\veta}=5$. The mean values of the metrics are reported in \cref{tab:6}, confirming a general behaviour comparable to the previous experiments. 
However, we point out that RELD largely outperforms the other diffusion-based approaches in terms of PSNR.   

\begin{table}
\centering
\caption{Values of PSNR, NIQE, PIQE and LPIPS for the images in Set5 corrupted setting $\downfact=4$, $\sigma_{\vA}=1$ and $\sigma_{\veta}=5$. Best results are highlighted in bold. }
\begin{tabular}{|c|c|c|c|c|c|}
\hline
& RED  & PNP & DPS & DiffPIR & RELD\\
\hline
PSNR $\uparrow$ & 23.81  & \textbf{25.23} & 20.81 & 21.50 & 23.32  \\
NIQE $\downarrow$ & 7.25  & 7.54 & 5.15 & 7.96 & \textbf{4.81} \\
PIQE $\downarrow$ & 83.88  & 86.66  & 50.50 & 44.72 & \textbf{42.32}\\
LPIPS $\downarrow$ & 0.360  & 0.332 & 0.350 & \textbf{0.298} & 0.303 \\

\hline
\end{tabular}
\label{tab:6}
\end{table}

We depict in \cref{fig:6} the results on an image in Set5 to visually inspect the performance of our model. It is evident that the proposed RELD is able to restore the finer details, as highlighted in the close-ups, which appear completely suppressed by the competing methods. 

\input{images/figure_table4_head_cut}
 
In conclusion, our results suggest that our proposal, RELD, achieves performances comparable with many state-of-the-art imaging techniques in terms of distortion quality. In particular, the numerical experiments prove that RELD produces more natural image reconstructions, as confirmed by the perceptual quality metrics like NIQE, PIQE and LPIPS.

\section{Conclusion} \label{sec:concl}

In this work, we presented the novel approach, Regularization by Latent Denoising (RELD), which combines latent denoising models with a Half-Quadratic Splitting optimization method. The main ingredients of this approach consist in operating in the latent space via a LDM trained specifically for a denoising task as suggested by Plug-and-Play and RED frameworks, in updating the novel iterates via HQS, in encompassing the embedding of $\vb$ in the input of the denoising network and in the joint optimization of latent and conditioning variables. The combination of these  strategies reduces the computational burden while still achieving high-quality results. The computational experience shows that RELD is competitive with state-of-the-art algorithms, providing more natural image reconstructions in terms of perceptual quality. 

Future work may comprise a theoretical foundation of the proposed method and insights on the error induced by employing a single gradient step in \cref{al:RELD}, instead of iteratively solving the optimization subproblem in $\vv$.

\section*{Acknowledgments}
A. Sebastiani is supported by the PRIN P2022J9SNP project “Advanced optimization METhods for automated
central veIn Sign detection in multiple sclerosis from magneTic resonAnce imaging  (AMETISTA)" project code: P2022J9SNP
(CUP E53D2301798001). A. Benfenati is supported by the Italian MUR through the PRIN 2022 Project "Sustainable Tomographic Imaging with Learning and rEgularization (STILE)", project code: 20225STXSB (CUP E53D23005480006).\\
This work is partially supported by the GNCS projects "Deep Variational Learning: un approccio combinato per la ricostruzione di immagini", "MOdelli e MEtodi Numerici per il Trattamento delle Immagini (MOMENTI)" and "Metodi avanzati di ottimizzazione stocastica per la risoluzion problemi inversi di imaging", projects codes: CUP E53C23001670001 and CUP E53C24001950001. A. Benfenati and A. Sebastiani are members of INdAM-GNCS.

\bibliographystyle{IEEEtran}
\bibliography{biblio.bib}
\end{document}